\documentclass[12pt]{iopart}
\usepackage{graphicx}
\usepackage{hyperref}
\hypersetup{
    unicode=false,          
    pdftoolbar=true,        
    pdfmenubar=true,        
    pdffitwindow=false,     
    pdfstartview={FitH},    
    pdftitle={Gravitational Radiation from CCSN},    
    pdfauthor={Konstantin N Yakunin},     
    pdfsubject={Subject},   
    pdfcreator={Creator},   
    pdfproducer={Producer}, 
    pdfkeywords={supernovae: general - numerical simulations - hydrodynamics - 
                          equation of state - neutrinos - gravitational waves}, 
    pdfnewwindow=true,      
    colorlinks=true,       
    linkcolor=red,          
    citecolor=blue,        
    filecolor=magenta,      
    urlcolor=cyan           
}

\newcommand\ba{\begin{eqnarray}}
\newcommand\ea{\end{eqnarray}}

\newcommand{\AAJ}{{\it Astron. Astrophys. }}

\begin{document}

\title[CCSN with CHIMERA: Gravitational Radiation]
       {Core Collapse Supernovae using CHIMERA: Gravitational Radiation from Non-Rotating Progenitors}

\author{Konstantin N Yakunin$^1$, 
Pedro Marronetti$^1$, 
Anthony Mezzacappa$^{2,3}$, 
Stephen W Bruenn$^1$, 
Ching-Tsai Lee$^3$, 
Merek A Chertkow$^3$, 
W Raphael Hix$^{2,3}$, 
John M Blondin$^4$, 
Eric J Lentz$^{2,3}$, 
O E Bronson Messer$^5$, 
and Shin'ichirou Yoshida$^6$}

\address{$^1$ Physics Department, Florida Atlantic University, Boca Raton, FL 33431-0991}
\address{$^2$ Physics Division, Oak Ridge National Laboratory, Oak Ridge, TN 37831-6354}
\address{$^3$ Department of Physics and Astronomy, University of Tennessee, Knoxville, TN 37996-1200} 
\address{$^4$ Department of Physics, North Carolina State University, Raleigh, NC 27695-8202}
\address{$^5$ National Center for Computational Sciences, Oak Ridge National Laboratory, Oak Ridge, TN 37831-6354}
\address{$^6$ Department of Earth Science and Astronomy, University of Tokyo}

\ead{cyakunin@fau.edu, pmarrone@fau.edu} 

\begin{abstract}
The CHIMERA code is a multi-dimensional multi-physics engine dedicated primarily to the simulation of core collapse supernova explosions. One of the most important aspects of these explosions is their capacity to produce gravitational radiation that is detectable by Earth-based laser-interferometric gravitational wave observatories such as LIGO and VIRGO. We present here preliminary gravitational signatures of two-dimensional models with non-rotating progenitors. These simulations exhibit explosions, which are followed for more than half a second after stellar core bounce.
\end{abstract}



\section{Introduction}
Of all the generators of Earth-detectable gravitational waves, core collapse supernovae (CCSN) are among the most interesting ones. The waveforms produced by these explosions will produce signals well within Advanced LIGO's \cite{AdvLIGO} sensitivity across a broad range of frequencies, carrying information about a number of phenomena, such as fluid instabilities in the proto-neutron star, neutrino-driven convection beneath the supernova shock wave, the standing accretion shock instability (SASI), deceleration at an aspherical shock, and aspherical neutrino emission (see Ott's review \cite{Ott2009}). This information is invaluable given that gravitational radiation, together with neutrino signals, are the only carriers of information that penetrate the heavy material expelled by a SN explosion.
We present here preliminary results from two-dimensional simulations carried out with the code CHIMERA, with emphasis in gravitational waveform (GW) generation.

\section{Two-Dimensional Simulations with CHIMERA}

The CHIMERA code is composed of different modules that handle the hydrodynamics, neutrino transport, self-gravity, a nuclear equation of state, and a nuclear reaction network. Details of these are given in \cite{Messer08, Bruenn09, Yakunin10}. Of special interest in this article is the gravitational physics module, which consists of a spectral Poisson solver used to determine the gravitational field \cite{Muller95} with general relativistic corrections to the spherical component \cite{Marek06}. The gravitational waves produced by CCSN are due to fluid dynamics and neutrino radiation. The details of the GW extraction methods employed in the post-processing analysis of CHIMERA data are given in \cite{Yakunin10}.

\begin{figure}
\begin{center}
\includegraphics[scale=0.8]{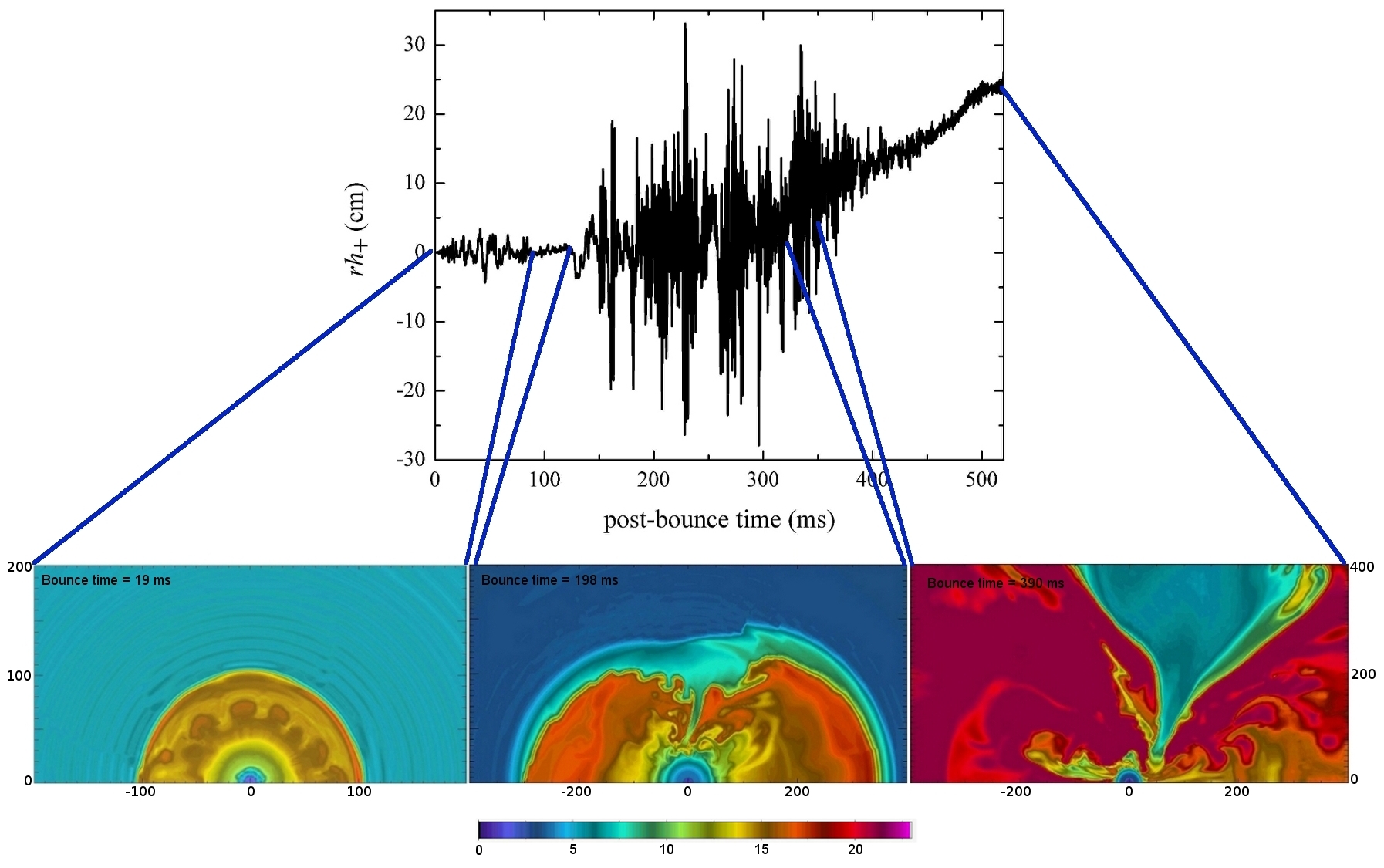}
\caption{{\it Left:} Gravitational wave strain $h_+$ times the distance to the observer $r$ vs. post-bounce time
for the 15 $M_{\odot}$ non-rotating progenitor
model. Below, entropy distribution snapshots typical of the {\it prompt}, {\it strong}, and {\it tail} stages of the signal.
Note the difference in scale of the left snapshot and two others.}
\label{figure1}
\end{center}
\end{figure}

The results presented here are based on axisymmetric simulations of CCSN for non-rotating 12, 15, and 25 $M_\odot$ progenitors \cite{Woosley07}. The gravitational signals produced by neutrino radiation and matter motion were followed for up to 530 ms after core bounce. The GWs are quantitatively different for every model but all of them show the same qualitative four stages of development:
\begin{itemize}
\item A {\bf prompt signal}: an initial and relatively weak signal that starts at bounce and ends at between 50 and 75 ms post-bounce.
\item A {\bf quiescent stage} that immediately follows the prompt signal and ends somewhere between 125 ms and 175 ms after bounce.
\item A {\bf strong signal}, which follows the quiescent stage and is the most energetic part of the GW signal.
 This stage ends somewhere between 350 ms and 450 ms after bounce. 
\item A {\bf tail}, which starts before the end of the strong signal at about 300 ms 
after bounce and consists of a slow increase in $rh_{+}$. This tail continues to rise at the end of our runs.
\end{itemize}

Gravitational waveforms for core collapse supernova have been obtained by other groups: Marek \etal \cite{Marek09}
showed the presence of the first three phases (prior to explosion) and Murphy \etal \cite{Murphy09} presented the same four stages based on parameterized explosions. Since our models are non-parameterized, the GWs presented here allow us to determine in a more precise manner the amplitudes and timescales. 

Figure \ref{figure1} shows a typical GW (this one corresponding to the 15 $M_\odot$ progenitor) together with snapshots of the entropy distribution corresponding to the three active stages of the signal. 

The first stage is composed of the contribution of two signals generated in different places: a high-frequency component produced inside the proto-neutron star (PNS), which has a radius of less than 30 km, and a low-frequency component originating at the shock radius (located at about 100 km at this time). The latter is due to the deflection of infalling matter through the shock, determined through the use of tracer particles \cite{Yakunin10}. Previous work attributed the prompt signal to PNS convection only \cite{Marek09, Murphy09}. 

The second stage is a period of relative calm that precedes the stronger phase of the signal: a third stage dominated by neutrino-driven convection and the Standing Shock Accretion Instability (SASI). During this stage, low-entropy downflows impinge on the PNS surface, creating the spikes shown in Fig. \ref{figure1}. The modulation of these funnels is driven by the SASI, affecting their kinetic energy and, consequently, the amplitude of the GWs generated when these flows hit the PNS.

The models end with GW tails of positive slope. These tails are associated with matter deflected at the expanding shock (see Fig. 5 in \cite{Murphy09}), characteristic of an explosion and are consistent with the prolate nature of the shock in the final stages of the simulations. (An oblate shock would give rise to a tail of opposite sign.)

\section{Conclusions and Future Work}

Our results mark  step forward in GW astrophysics, but it is important to remark on their preliminary nature: we are currently performing a new set of runs using an enhanced version of CHIMERA that will provide an essential test of the validity of the GWs presented here \cite{Bruenn10}. Moreover, the most important limitation in these models is their restriction to axisymmetry. We are currently evolving three-dimensional models, and we anticipate that the greatest change to our GW predictions will be in the phase-4 tail. This is because
prolate explosions are usually present in axisymmetric simulations, where artificial boundary conditions must be imposed that prevent the turnover of material along the symmetry axis. Additionally, the amplitude and timescales associated with the early stages are bound to change too, since three-dimensional simulations of the SASI have shown spiraling flows that cannot be captured by our current 2D models  \cite{Blondin07}.

\section{Acknowledgements}

The authors would like to acknowledge the computational resources provided at the Leadership Computing Facility 
in the National Center for Computational Sciences at ORNL (INCITE Program) and at TACC (TG-MCA08X010). 
PM acknowledges partial support 
from NSF-PHYS-0855315, and PM and SWB acknowledge partial support from an NSF-OCI-0749204 award. AM, OEBM, 
PM, SWB, and WRH acknowledge partial support from a NASA ATFP award (07-ATFP07-0011). AM and WRH acknowledge 
support from the Office of Nuclear Physics, U.S. Department of Energy, and AM and OEBM acknowledge support from the Office 
of Advanced Scientific Computing Research, U.S. Department of Energy.

\end{document}